\documentclass[twocolumn,prl,superscriptaddress,longbibliography%,nofootinbib
]{revtex4-1}
\usepackage{graphicx}
\usepackage{amssymb}
\usepackage{amsmath}
\usepackage{color}
\usepackage{comment}
\usepackage{bm}

\newcommand{\up}{\uparrow}
\newcommand{\down}{\downarrow}

\newcommand{\ket}[1]{\left|{#1}\right>}

\newcommand{\beq}{\begin{equation}}
\newcommand{\eeq}{\end{equation}}

\usepackage{hyperref}
\hypersetup{%
   pdfpagemode=None, %FullScreen,
   pdfstartpage=1,
   pdfmenubar=true,
   pdftoolbar=true,
   colorlinks = true,
   linkcolor=blue,
   citecolor=blue,
   urlcolor=blue,
   bookmarksopen=false
 }  

\begin{document}

\title{Magnetism in strongly interacting one-dimensional quantum mixtures}

\author{Pietro Massignan}
\affiliation{ICFO-Institut de Ciencies Fotoniques, The Barcelona Institute of Science and Technology, 08860 Castelldefels (Barcelona), Spain}

\author{Jesper Levinsen}
\email{jesper.levinsen@monash.edu}
\affiliation{School of Physics and Astronomy, Monash University, Victoria 3800, Australia.}

\author{Meera M. Parish}
\affiliation{School of Physics and Astronomy, Monash University, Victoria 3800, Australia.}

\date{\today}

\begin{abstract}
  We consider two species of bosons in one dimension near the
  Tonks-Girardeau limit of infinite interactions. For the case of equal masses and
  equal intraspecies interactions, the system can be mapped to a
  $S=1/2$ XXZ Heisenberg spin chain, thus allowing one to access
  different magnetic phases. 
  Using a powerful ansatz developed for the
  two-component Fermi system, we elucidate the evolution from few to
  many particles for the experimentally relevant case of an external
  harmonic confinement.  In the few-body limit, we already find clear
  evidence of both ferromagnetic and antiferromagnetic spin
  correlations as the ratio of intraspecies and interspecies
  interactions is varied.  Furthermore, we observe the rapid emergence of symmetry-broken
  magnetic ground states as the particle number is increased. 
 We therefore demonstrate that systems containing only a few bosons are
    an ideal setting in which to realize the highly sought-after itinerant ferromagnetic phase.
\end{abstract}


\maketitle

Quantum magnetism is ubiquitous in nature and plays a central role in
important phenomena such as high-temperature superconductivity
\cite{Monthoux2007}. Furthermore, it underpins the technological
advances in data storage \cite{GMR}, and it promises a new generation
of spintronic devices, where the spin of the electron rather than its
charge is used to transfer information. However, despite its ubiquity,
magnetic phenomena are often difficult to characterise and treat
theoretically; for instance, itinerant ferromagnetism of delocalized
fermions requires strong interactions and is thus not completely
understood \cite{Moriya}.

One can gain insight into magnetic phases by considering cleaner, more
idealised versions of the phenomena.  In particular, an important
question is whether or not ferromagnetism can exist \emph{without an
  underlying lattice}~\cite{Stoner1939}.  This possibility was
investigated experimentally in atomic Fermi gases
\cite{Ketterle2009,Ketterle2012}, but the system proved to be unstable
towards fermion pairing rather than magnetism.  More recently, it has
been proposed that itinerant ferromagnetism can be realised in a
strongly interacting one-dimensional (1D) Fermi
gas~\cite{cui2013}. However, in this case, one cannot access the
ferromagnetic phase without explicitly breaking the spin symmetry with
an external field~\cite{Cui2014}, and this potentially complicates the
observation of ferromagnetism.

\begin{figure}
\centering
\vskip 0 pt
\includegraphics[width=0.96\columnwidth,clip]{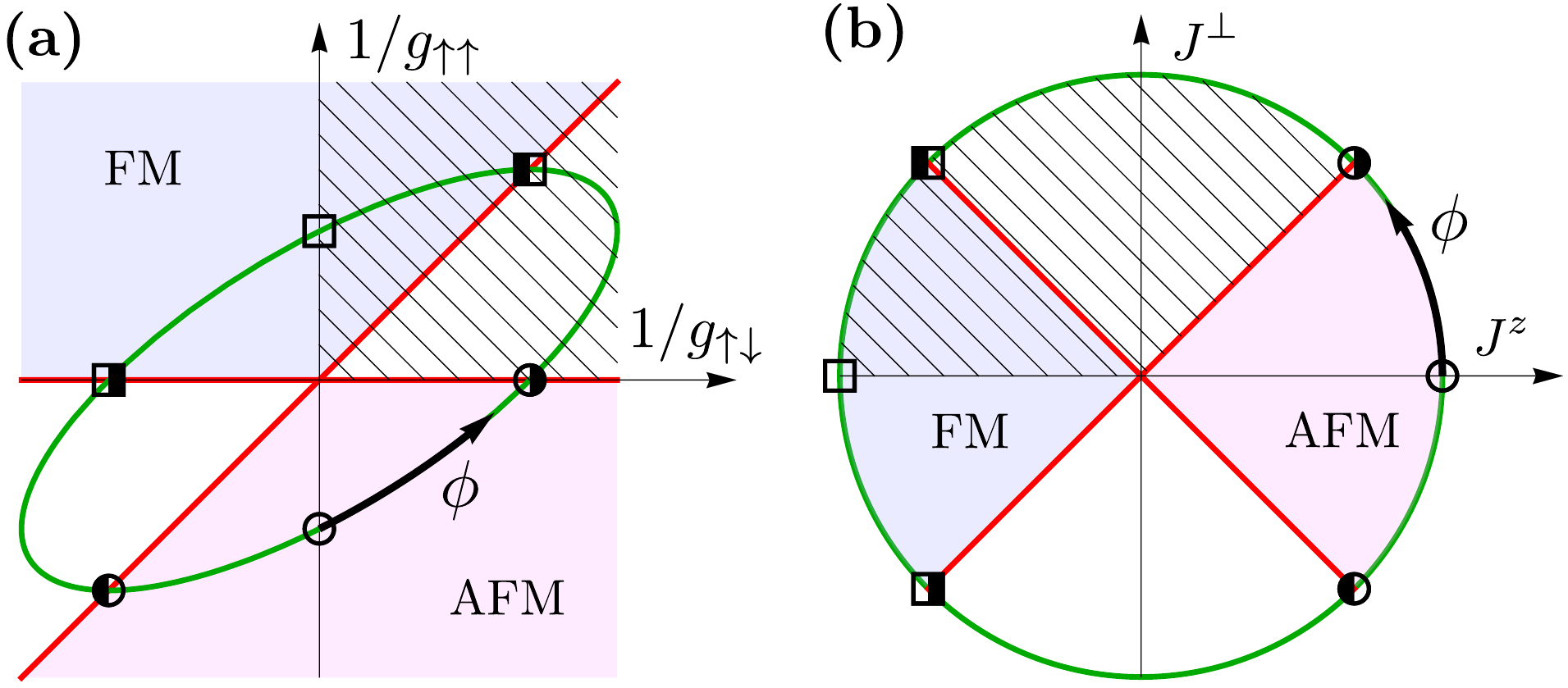}
%\hspace{.1cm}
%\includegraphics[width=0.48\columnwidth,clip]{fig_sketchvsphi.pdf}
\caption{(a) Magnetic phase diagram for the %\pietro{remove? trapped} 
two-component Bose
  gas near the Tonks-Girardeau limit of infinite interactions,
  $1/g_{\nu\nu'} \to 0$, where we focus on the ``upper-branch'' ground
  state and ignore any two-body bound states.
  (b)  The system may be
  mapped onto an XXZ Heisenberg spin model, Eq.~\eqref{eq:spinH}, with
  couplings $J^\perp$ and $J^z$ in the $x$-$y$ plane and $z$
  direction, respectively.  
  Equivalent points in the phase diagrams are marked by identical symbols, 
  e.g., $\phi=3\pi/4$ corresponds to identical inter- and intraspecies interactions.
  The thick (red) lines correspond to the
  XXX Heisenberg model that is realized by the two-component Fermi
  gas.
In the thermodynamic limit, these lines bound regions of Ising ferromagnetism (FM) and Ising antiferromagnetism (AFM), separated by disordered ``spin liquid" phases.
  In the hatched region all interactions are repulsive, and the spin-chain %upper-branch
  ground state is the actual ground state of the system.  
}
\label{fig:sketch}
\end{figure}

In this Letter, we show that both itinerant ferromagnetism and
antiferromagnetism can be investigated with a 1D two-component mixture
of bosons \footnote{Here, we mean ``itinerant'' in the sense that the
  particles are not localized by an underlying lattice.}.  For the
case of equal masses and equal intraspecies interactions
($g_{\up\up}=g_{\down\down}$), the Bose-Bose mixture may be regarded
as a pseudo-spin $S = 1/2$ system, and it can be mapped to a 1D spin
chain in the limit of strong interactions~\cite{Guan2007,Matveev2008}.  In
particular, for infinite intraspecies interactions, the system is
formally equivalent to a 1D Fermi gas. However, in contrast to the
Fermi case, one can break the $SU(2)$ symmetry and access different
magnetic phases by varying the ratio of intraspecies and interspecies
interactions (see Fig.~\ref{fig:sketch}). Moreover, the
ferromagnetic state should be stable, unlike in higher
dimensions~\cite{Pekker2011,Massignan2014,Ngampruetikorn2012}.

Experimentally, 1D quantum gases have been successfully realized with
strongly interacting bosons~\cite{Paredes2004,Kinoshita2004}, two
species of fermions~\cite{Liao2010,Serwane2011,Zurn2012,Wenz2013},
and, more recently, fermions with $SU(n)$ symmetry, where the number
of spin components can be tuned from $n=2$ to 6 \cite{Pagano2014}.
All these 1D experiments feature an underlying harmonic trapping
potential, which has enabled the study of the  evolution from few
to many particles~\cite{Serwane2011,Wenz2013,Grining2015}. However,
it is theoretically challenging to treat
1D interacting particles in a harmonic trap, since the problem cannot be solved
exactly in general \cite{Guan2013}. Only a small number of trapped
particles can be treated exactly
numerically~\cite{Busch1998,Gharashi2013,Sowinski2013,Astrakharchik2013,Deuretzbacher2014,Volosniev2014,GarciaMarch2014,Gharashi2015,Dehkharghani2015,Grining2015}.

Recently, we developed a highly accurate ansatz for the strongly
interacting 1D Fermi gas in a harmonic potential that allowed us to
analytically determine the wavefunction for a single
spin-$\down$ impurity with any number of majority $\up$ fermions
\cite{Levinsen2015}.  Here, we demonstrate that this approach can be
extended to provide an extremely accurate description of the Bose
mixture, providing a benchmark for other numerical techniques. 
We use this method to address the question of how the ground state of the few-body system is connected to magnetic phases in the many-body limit.
As the particle number $N$ is increased, we observe the rapid emergence of a
doubly degenerate ground state, corresponding to the $Z_2$ symmetry
that is ultimately broken within the Ising magnetic phases in the
thermodynamic limit.  
We furthermore argue that stable ferromagnetism
is best observed in a 1D \emph{few-body} system.

\textit{Effective spin model.---} 
We start by considering a 1D two-component ($\nu= \up , \down$)
gas of bosons with mass $m$, trapped in a harmonic potential with
frequency $\omega$. Inter- and intraspecies interactions are assumed
to be short-ranged, and parametrized respectively by $g_{\up\down}$
and $g_{\nu\nu}$.  The Hamiltonian is
\begin{align} \notag
  H = &
  \sum_{\nu} \int dx \ \psi_\nu^\dag(x) \left[-\frac{\hbar^2}
{2m}\frac{\partial^2}{\partial x^2} +\frac{1}{2} 
m\omega^2x^2\right]  \psi_\nu(x) \\ 
  & + \frac{1}{2}\sum_{\nu,\nu'} g_{\nu\nu'} \int d x 
\ \psi^\dag_\nu(x) \psi^\dag_{\nu'}(x) \psi_{\nu'}(x) \psi_\nu(x),
\label{eq:H}
\end{align}
where $\psi_\nu^\dag$ and $\psi_\nu$, respectively, create and
annihilate a boson with pseudo spin $\nu$.  In the following we use
harmonic oscillator units, where $\hbar=m=\omega=1$. Moreover, we
consider the symmetric case where the two intraspecies
interactions are of equal strength and the number of particles in each
spin state is the same, i.e., $g_{\up\up}=g_{\down\down}$ and
$N_\up=N_\down\equiv N/2$.

In the Tonks-Girardeau limit $|g_{\nu\nu'}|\to\infty$,
the particles
become impenetrable and
retain their particular ordering.
Thus, even in the absence of an underlying lattice, one can consider the system as a discrete
chain
of length $N$, where the site index defines the position along
the ordered chain. 
Perturbing away from this limit allows nearest neighbors to exchange
position and one can describe the system using a spin-chain model
\cite{Matveev2004,Matveev2008,Volosniev2014,Deuretzbacher2014,Volosniev2015,Levinsen2015}. 
To make this connection evident, 
it is convenient to replace the inter- and intraspecies interactions $g_{\up\down}$ and $g_{\up\up}=g_{\down\down}$
by effective parameters $g>0$ and $\phi$, such that the entire phase
diagram is traversed by changing the ``angle'' $\phi$, as depicted in 
Fig.~\ref{fig:sketch}.
 These are related to the original interactions by
$\sin\phi = g/g_{\up\down}$ and
$\cos\phi = g/g_{\up\down} - 2g/g_{\up\up}$.  
To linear order in $1/g$, the original Hamiltonian may then be written as  \cite{McCulloch2014,Volosniev2015}:
\begin{align}
H=\varepsilon_0+(\Delta \varepsilon+{\cal H})/g.
\label{eq:H2}
\end{align}
Here, $\varepsilon_0$ is the energy of $N$ identical fermions, and ${\cal H}$ is
an effective Heisenberg XXZ
 Hamiltonian:
\begin{align}
& {\cal H} =  \sum_{i=1}^{N-1}  \eta_i
\left[J^\perp\left(\sigma_i^x \sigma_{i+1}^x +  \sigma_i^y
  \sigma_{i+1}^y \right)+ 
J^z \sigma_i^z\sigma_{i+1}^z \right],
\label{eq:spinH}
\end{align}
where 
$J^\perp \equiv \sin\phi$, $J^z \equiv \cos\phi$, and ${\boldsymbol\sigma}_i$ are the
Pauli spin matrices at position $i$.
The constant
$\Delta \varepsilon =(J^z- 2J^\perp) \sum_{i=1}^{N-1} \eta_i$ ensures
that only identical (distinguishable) bosons interact via $g_{\up\up}$
($g_{\up\down}$).
Note that, apart from parity, the properties of the
  eigenstates of ${\cal H}$  are invariant under the transformation $\phi\mapsto-\phi$~\cite{supmat}.

 The energy cost of exchanging
opposite spins at positions $i$ and $i+1$ depends on the underlying
external confinement, an effect encapsulated in the coefficients
$\eta_i$ of Eq.~\eqref{eq:spinH}.
These must in general be calculated numerically,
but a key simplification is that they are independent of the
interactions and thus one need only determine them for a single %angle
$\phi$.  In particular, for infinite intraspecies interaction
($\phi = \pi/4$, $5\pi/4$), the system is equivalent to a
two-component Fermi gas, where the $\eta_i$ have already been
investigated for $N\leq9$ 
\cite{Deuretzbacher2014,Volosniev2014,Levinsen2015}.
We have recently shown for the
harmonically trapped Fermi gas that $\eta_i$ is well approximated by
the analytic expression $i(N-i)\bar\eta(N)$ \cite{Levinsen2015}.  
Referring to the Supplemental Material~\cite{supmat}, we find that this approach also yields
near exact wavefunctions for {\em any} $\phi$.
We will, thus, use this ansatz to consider larger numbers of particles.

\textit{Phase diagram.---}
To gain insight into the behavior of the trapped boson system,
it is instructive to first consider the uniform XXZ model ($\eta_i=1$),
where the magnetic phases are well characterized~\cite{magnetism,Imambekov2012}.
In the thermodynamic limit, 
the local density approximation (LDA) shows that the topology 
of the uniform phase diagram 
is the same as
that of the trapped system.
As displayed in Fig.~\ref{fig:sketch}, we expect to have fully saturated ferromagnetism (FM) of the Ising type 
when $J^z< - |J^\perp|$, i.e., when $3\pi/4 < \phi < 5\pi/4$ \cite{magnetism}. 
Indeed, the first-order transition at $J^z = -J^\perp < 0$ corresponds to $g_{\up\down} =  g_{\nu\nu}>0$, 
the criterion for phase separation in a weakly interacting two-component Bose gas~\cite{Imambekov2012}.
However, as we discuss below, the interface between $\up$ and $\down$ bosons is sharper in the strong-coupling limit, being of order the interparticle spacing.
The other first-order FM transition, $J^z = J^\perp<0$, has no weak-coupling analogue
%analogue at weak coupling
 since it occurs 
on the metastable ``upper branch'' for attractive interactions (or ``super-Tonks" regime).

When $|J^\perp| = |J^z|$,
${\cal H}$ becomes equivalent to the XXX Heisenberg model
that can be realized with fermions, e.g., $\phi = 3\pi/4$ can be
mapped to the upper branch of the attractive Fermi gas
\cite{Deuretzbacher2014,Volosniev2015}, which is the regime where
itinerant ferromagnetism exists~\cite{cui2013}.
Once $1/g_{\nu\nu} < 0$,
we will also have Ising antiferromagnetism (AFM)
for $-\pi/4 < \phi < \pi/4$, but the phase transition to this state is
of the Kosterlitz-Thouless type in the uniform case \cite{magnetism}.

\begin{figure}
\centering
\vskip 0 pt
\includegraphics[width=0.95\columnwidth,clip]{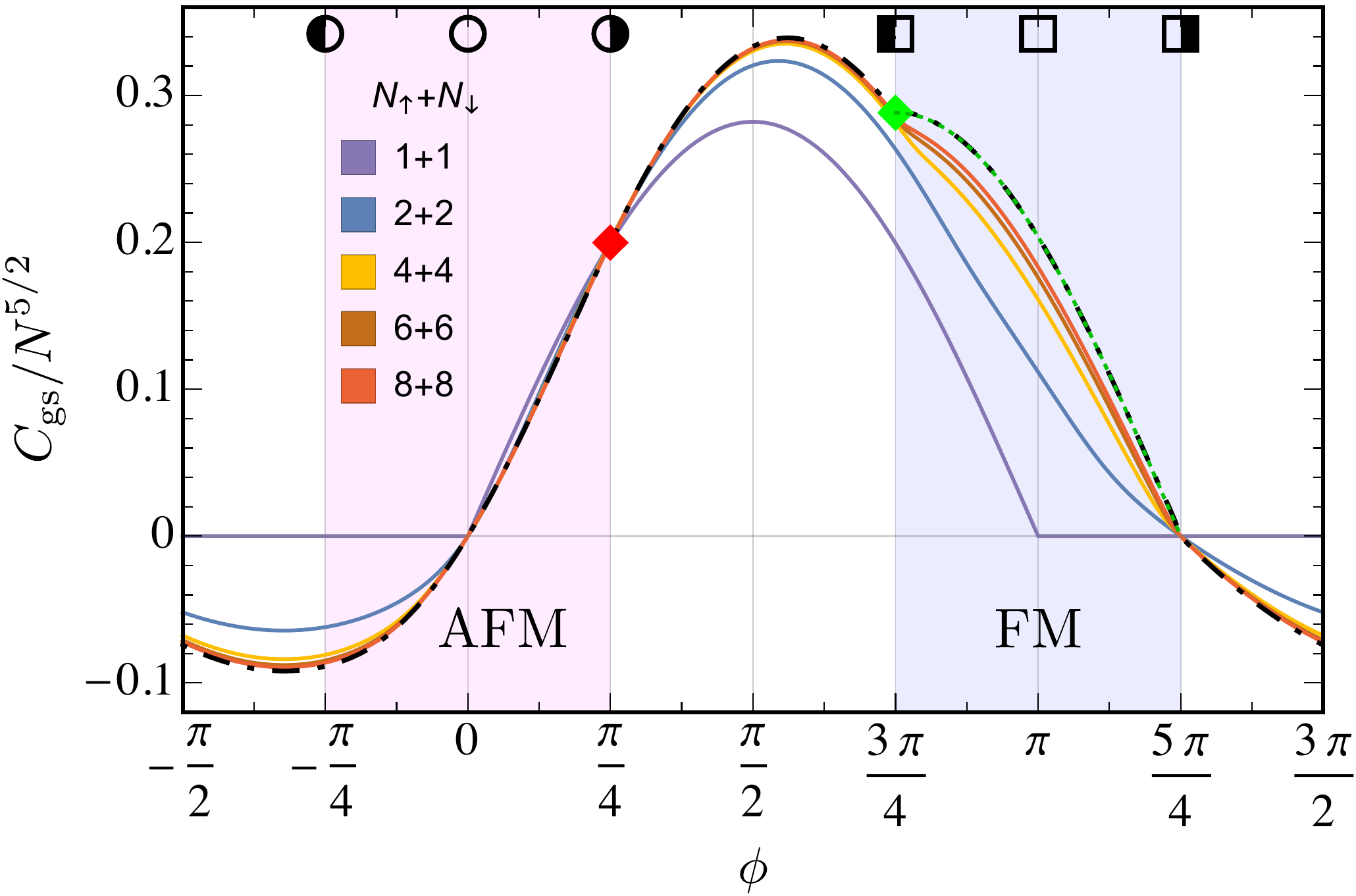}
\caption{Shift in the ground-state energy (see main text) as a
  function of $\phi$ for increasing particle number,
from $N=2$ to $16$ (solid lines).
The dash-dotted black line is an
  extrapolation to the limit $N \to \infty$ \cite{extrap}.  
  The black symbols and shaded
  regions correspond to the interaction strengths and magnetic phases illustrated in
  Fig.~\ref{fig:sketch}.
    Diamonds indicate analytical Bethe
  ansatz results %for $N$ particles 
within LDA, where $\phi = \pi/4$ corresponds to a two-component
  Fermi gas \eqref{GYcontact}, and $\phi = 3\pi/4$ is equivalent to
  identical bosons \eqref{LLcontact}.  The dotted green line
  corresponds to the result for spin-polarized bosons for $N\to\infty$.
  }
\label{fig:GScontact}
\end{figure}

\emph{Ground-state energy.---} 
We now turn to the question of how the few-body system approaches the many-body limit.
We first consider the ground-state
energy of the trapped gas:
$E_{0}=\varepsilon_0-C_{\rm gs}/g$ \cite{notecontact}.  Since
$\varepsilon_0$ is a simple energy shift, we take
$C_{\rm gs}$ as a function of $\phi$, as shown in
Fig.~\ref{fig:GScontact}.  The result for $N_\up=N_\down = 1$ exhibits
discontinuities in the slope at $\phi = 0$, $\pi$, corresponding to
level crossings in the energy spectrum, where the total spin of the
ground state is $S = 0$ for $0 < \phi < \pi$ and $S=1$ for
$-\pi < \phi < 0$. In this case, there is no intraspecies interaction
to couple states of different total spin, similarly to the
two-component Fermi gas, and one cannot adiabatically evolve the
ground state.  However, for $N>2$, the total spin is no longer
conserved and thus one can access different magnetic phases
by varying $\phi$ adiabatically.

Figure~\ref{fig:GScontact} shows that the results rapidly approach a universal curve 
with increasing $N$.
In particular, we see that certain $\phi$ extrapolate
to known Bethe ansatz results within LDA \cite{Astrakharchik2005}. For $\phi = \pi/4$, where our system
corresponds to a two-component Fermi gas,
we recover the LDA ground-state energy of the repulsive Gaudin-Yang model
\cite{Astrakharchik2004,Astrakharchik2005,ColomeTatche2008}:
\begin{align} \label{GYcontact}
\frac{E_{\rm GY}}{N^2} = \frac{1}{2} - \frac{128\sqrt{2}\ln(2)}{45\pi^2} \frac{\sqrt{N}}{g_{\up\down}} \> .
\end{align}
For $\phi = 3\pi/4$, we have
$g_{\nu\nu} = g_{\up\down}$ and the ground state is equivalent to that
of a strongly repulsive gas of $N$ identical bosons.  Thus,
we see that our results converge to the LDA
ground-state energy of the Lieb-Liniger model at strong coupling
\cite{Astrakharchik2005}:
\begin{align} \label{LLcontact}
\frac{E_{\rm LL}}{N^2} = \frac{1}{2} - \frac{128\sqrt{2}}{45\pi^2} \frac{\sqrt{N}}{g_{\up\up}}.
\end{align}

Within the entire region $3\pi/4 \leq \phi \leq 5\pi/4$, we find that the
ground-state energy in the limit $N \to \infty$ corresponds to that of
spin-polarized bosons, where it follows from
Eq.~\eqref{LLcontact} that
$C_{\rm gs} = \frac{128 N^{5/2}}{45\pi^2} \cos (\phi - 3\pi/4)$.
For fixed spin populations $N_\up = N_\down$, this result physically
corresponds to phase separation of $\up$ and $\down$ particles, where
we can neglect the energy of the interface in the thermodynamic limit
since it scales like $1/N$ compared to the total energy $E_0$.  The
formation of spin-polarized domains is a signature of 
ferromagnetism~\footnote{Note that a similar ``easy-axis'' FM phase is predicted to exist in 3D~\cite{Radic2014}. }, 
and the emergence of kinks in the energy at $\phi=3\pi/4$ and
$5\pi/4$ clearly indicates a first-order transition to the FM phase
(Fig.~\ref{fig:GScontact}).
While there are no sharp features in the energy at the AFM phase boundaries, 
the fact that $C_{\rm gs} = 0$ at $\phi = 0$ is a direct consequence
of classical N\'eel ordering, where identical spins are staggered and
thus experience no intraspecies interaction.

\begin{figure}[h]
\centering
\vskip 0 pt
\includegraphics[width=\columnwidth,clip]{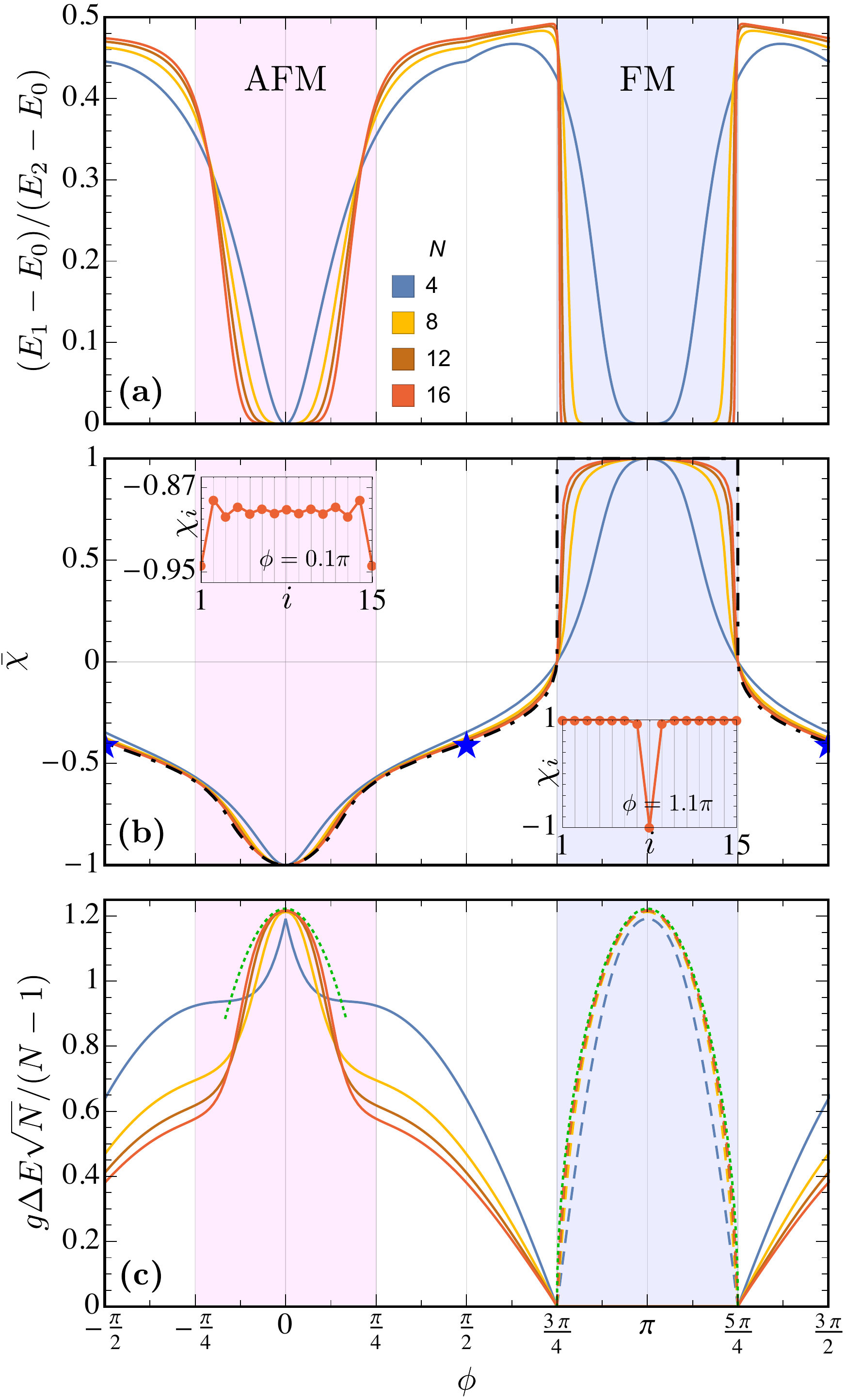}
\caption{(a) Energy-difference of ground and first
  excited states, $E_1-E_0$, compared to the energy of the second
  excited state, $E_2$.
(b) Trap-averaged magnetic correlation function evaluated in
  the ground state.  The dash-dotted line is an extrapolation to the
  thermodynamic limit \cite{extrap}, and the stars are
  analytic results for a uniform system.  The insets show the
  spatially-resolved correlation function for $N=16$ and two different
  values of $\phi$.
(c) Spin gap $\Delta E$ for the spin polarised (dashed) and unpolarised (solid) ground states.
The green dotted lines are the analytical results for large $N$ 
 --- see Supplemental Material \cite{supmat}.
    }
\label{fig:degeneracyAndCorrelationsAndSpinGaps}
\end{figure}

\emph{Magnetic phases.---} 
A clear signature of magnetic order can be extracted from the degeneracy of the ground
state. A two-fold degenerate ground state is a \emph{necessary}
condition for Ising magnetic order, and thus its emergence in the
few-body system is a precursor for symmetry-broken magnetic states as
$N \to \infty$.  As shown in Fig.\
\ref{fig:degeneracyAndCorrelationsAndSpinGaps}(a), the two
lowest energy states become rapidly degenerate with increasing $N$ in
the FM regime.  In the few-body system, these two states have opposite
parity, but as they become degenerate they can be combined to form
states that break spin symmetry and parity.  Physically, this
degeneracy emerges once the tunnelling between different symmetry
broken states is suppressed for large $N$.
A similar situation
occurs in the AFM region, but the degeneracy manifests more slowly.
Therefore, very few particles are necessary to observe strong magnetic
correlations and a large breaking of the $Z_2$ symmetry in the FM
region, while larger systems are needed to observe similar effects in
the AFM region.

We can further characterize the different phases using
the short-range correlation function
$\chi_i=\langle \sigma^z_{i}\sigma^z_{i+1}\rangle$ in the ground
state.  Referring to the insets of
Fig.~\ref{fig:degeneracyAndCorrelationsAndSpinGaps}(b), we see
that $\chi_i$ is always close to $-1$ around $\phi = 0$, signifying
AFM correlations. Likewise, for the FM phase, we obtain
$\chi_i \approx 1$ throughout the spin chain, excluding
the interface between the spin-polarized domains. For strong interactions, this domain
wall is sharp, decaying exponentially with distance even in the
few-body system.

To obtain a more global description of magnetic correlations, we take
the trap-averaged function
$\bar{\chi}(\phi)=\frac{1+\sum_i\chi_i}{N-2}$, normalized such
that
$\bar{\chi}(0) =-1$ and $\bar{\chi}(\pi) = 1$.
At the FM boundaries,
the ground state is a Heisenberg
ferromagnet with total spin $S=N/2$ oriented in the $x$-$y$ plane.
Here, the wavefunction is obtained by applying the total spin lowering
operator $N/2$ times to the spin-polarized system
$\ket{\up \up ... \up}$.  Thus, by symmetry,
$\bar{\chi} = \chi_i = 0$.  With increasing $N$, we find that the
averaged short-range correlations approach
 a rectangular function in the FM region, where $1 - \bar{\chi} \simeq (\pi - \phi)^2/N$ around
$\phi=\pi$ for $N \gg 1$.  This is consistent with the appearance of
classical Ising FM along the $z$ direction.

By contrast, we see that quantum fluctuations reduce the Ising AFM
correlations around $\phi = 0$, with $1+\bar{\chi} \simeq \phi^2$ in
the limit $N \to \infty$.  Furthermore, $\bar{\chi}$ evolves smoothly
with $\phi$ from the AFM phase to the disordered ``spin liquid''
occupying the region $\pi/4 < |\phi| < 3\pi/4$.  The disordered liquid
also features $\bar{\chi} < 0$ since its spin correlations resemble
those in a Fermi system.  Indeed, for $\phi = \pm \pi/2$, the spin
model can be formally mapped onto free fermions using the
Jordan-Wigner transformation \cite{Imambekov2012}.  For the uniform
case, the free-fermion problem can be solved analytically and we find
$\bar{\chi} = -4/\pi^2$, which compares well with the large $N$ limit
of the trapped system
[see Fig.~\ref{fig:degeneracyAndCorrelationsAndSpinGaps}(b)].

\emph{Spin imbalance.---} 
When $N_\up \neq N_\down$, magnetic correlations strongly influence the behavior of the trapped system.
In the limit of a single impurity $N_\down =1$, we find that the type of correlations determines the position of the $\down$ impurity: in the AFM regime, the impurity sits at the trap center~\cite{Levinsen2015}, while FM correlations confine it to the edge.
More generally, we find in the few-body system that the fully spin-polarised state always has the lowest energy within the FM regime, as expected, while the ground state for all other $\phi$ is unpolarised. 
In practice, one can observe this by adding 
a small spin-interconversion coupling~\cite{McCulloch2014}.
To quantify this behavior, we determine 
the energy cost of flipping a spin in the
ground state [Fig.~\ref{fig:degeneracyAndCorrelationsAndSpinGaps}(c)].
For the fully polarized FM ground state, we see that the ``spin gap'' $\Delta E$ always vanishes at the FM phase boundaries and
rapidly converges to a universal curve with increasing $N$ \cite{supmat}.
This further supports the idea that FM can be realized in a few-boson system.
On the other hand, for the unpolarized system we find that the AFM spin
gap remains finite even in the disordered phase, illustrating the
different nature of the spin excitations. 
In both cases, we find that the behavior in the trap is qualitatively different
from that in the  uniform system, since the flipped spins predominantly reside
at the trap edges and, hence, the spin gaps may not be obtained via the
LDA.

\emph{Concluding remarks.---}
The physics described in this Letter may be realized in a
two-component Bose gas close to overlapping 1D inter- and intraspecies
resonances, which obviously necessitates some fine tuning. However, we
emphasize that the 1D Bose gas is substantially more tunable than its
3D counterpart: Firstly, the magnetic field need not be set exactly to
any of the 3D resonances, since part of the fine tuning can be done by
varying the strength of the transverse confinement
\cite{Olshanii1998}, and this may even be done in a species-selective
manner. Secondly, we propose to employ {\em any} two hyperfine spin
states, as spin-changing collisions will be suppressed in the strongly
interacting ``fermionized'' limit.  Thus we anticipate that
the strongly-coupled two-component Bose gas could be investigated in
current experimental setups.

We have argued that clear signatures of a first-order quantum phase
transition to a ferromagnetic state -- the analogue of itinerant
electron Stoner ferromagnetism -- appear already in the few-body
limit. Furthermore, we emphasize that the few-body scenario is likely to be ideal for observing
magnetic phases since the condition of staying in the strongly
interacting regime, $g/\sqrt{N}\gg1$, becomes increasingly difficult
to satisfy as $N$ increases. 
Smaller values of  $N$ also enlarge the relative energy spacing
between eigenstates, thus making it easier to tune interactions
adiabatically and access stable magnetic ground states that are robust with respect to thermal fluctuations.
Hence, the harmonically trapped and
strongly interacting few-body Bose gas appears ideally suited for the realisation of quantum magnetic phases.

%%%%%%%%%%%%%%%%%%%%%%%%%%%%%%%%%%%%%%%%%%%%%%%%%

\begin{acknowledgments}
{\bf Acknowledgments:} We gratefully acknowledge fruitful discussions
with Victor Galitski, Xi-Wen Guan, Leticia Tarruell, Lincoln Turner,
and Artem Volosniev.  P.M. acknowledges support from ERC AdG OSYRIS,
EU EQuaM, Plan Nacional FOQUS, 2014 SGR 874, Spanish MINECO (Severo Ochoa
grant SEV-2015-0522), and the Ram\'on y Cajal
programme.
\end{acknowledgments}

%\bibliography{1DBose}

\begin{thebibliography}{49}
\bibitem [{ext()}]{extrap}%
  \BibitemOpen
  \href@noop {} {}\bibinfo {note} {At each value of $\phi$ in Fig.\ 2 [3(a)],
  we extrapolate to $N=\infty$ by fitting a quadratic polynomial to the
  few-body results for $N=4,6,8,10,12,14,16$ vs.\ $1/N$ [$1/N^2$].}\BibitemShut
  {Stop}%
\bibitem [{not()}]{notecontact}%
  \BibitemOpen
  \href@noop {} {}\bibinfo {note} {$C_{\rm gs}$ is proportional to the usual
  contact~\cite{Tan2008,Tan2008b} when the intraspecies interactions are
  infinite or when $g_{\up\up} = g_{\up\down}$.}\BibitemShut {Stop}%
\end{thebibliography}


%%%%%%%%%% Merge with supplemental materials %%%%%%%%%%
\widetext
\clearpage
\begin{center}
\textbf{\large Supplemental Material:
			Magnetism in strongly interacting 1D quantum mixtures}\\
\vspace{4mm}
{Pietro Massignan,$^1$ Jesper~Levinsen,$^2$ and Meera~M.~Parish,$^{2}$}\\
\vspace{2mm}
{\em \small
$^1$
ICFO-Institut de Ciencies Fotoniques, Barcelona Institute of Science and Technology, 08860 Castelldefels (Barcelona), Spain\\
$^2$School of Physics and Astronomy, Monash University, Victoria 3800, Australia\\
}\end{center}
%%%%%%%%%% Merge with supplemental materials %%%%%%%%%%
%%%%%%%%%% Prefix a "S" to all equations, figures, tables and reset the counter %%%%%%%%%%
\setcounter{equation}{0}
\setcounter{figure}{0}
\setcounter{table}{0}
\setcounter{page}{1}
\makeatletter
\renewcommand{\theequation}{S\arabic{equation}}
\renewcommand{\thefigure}{S\arabic{figure}}
\renewcommand{\thetable}{S\arabic{table}}
%\renewcommand{\bibnumfmt}[1]{[S#1]}
%\renewcommand{\citenumfont}[1]{S#1}
%%%%%%%%%% Prefix a "S" to all equations, figures, tables and reset the counter %%%%%%%%%%

\section{Symmetries of the Hamiltonian}

As discussed in the manuscript, when we are in the Tonks-Girardeau limit $|g_{\nu\nu'}|\to\infty$, the system may be described in terms of an effective XXZ Hamiltonian, 
\begin{align}
& {\cal H} =  \sum_{i=1}^{N-1}  \eta_i
\left[J^\perp\left(\sigma_i^x \sigma_{i+1}^x +  \sigma_i^y
  \sigma_{i+1}^y \right)+ 
J^z \sigma_i^z\sigma_{i+1}^z \right],
\label{eq:spinHsupp}
\end{align} 
where the parameters are defined as in the main text.
For even particle number $N$, as is the case when the spin populations are balanced ($N_\up=N_\down$), one can show that this spin-chain
Hamiltonian satisfies
$\mathcal{H}(-\phi) = U \mathcal{H}(\phi) U^\dag$, with the unitary operator $U \equiv
\prod_i^{N/2} \sigma_{2i}^z$.  Thus, with the exception of parity, we expect the
properties of the eigenstates to be
invariant under $\phi \mapsto -\phi$. The parity operator $P\equiv\prod_{i=1}^N\sigma_i^x$, 
so that for even $N/2$ we have $[U,P]=0$ and the ground state always has the
same parity. For odd $N/2$, instead $\{U,P\}=0$, and 
consequently the ground state changes parity at $\phi=0$ and $\pi$.
On the other
hand, the total energy is not invariant under the unitary
transformation due to the presence of the constant
$\Delta \varepsilon (\phi)$ in the strong-coupling Hamiltonian, Eq.~\eqref{eq:H2}.

\section{Parameters of the model}

The parameters $\eta_i$ which yield the values of the spin-exchange
coefficients of the spin-chain Hamiltonian \eqref{eq:spinHsupp} % in the manuscript 
were computed in Refs.~\cite{Volosniev2014,Levinsen2015,Deuretzbacher2014}. For
completeness, we list these in Table~\ref{tab:eta} up to $N=9$.

\begin{table}[h]
\begin{minipage}{\textwidth}
\begin{tabular}{|c|c|c|c|c|c|c|c|c|}
\hline
$N$ & 2 & 3 & 4 & 5 & 6 & 7 & 8 & 9 \\
\hline
$\eta_1$ & $\sqrt{1/2\pi}$ & $27/(16\sqrt{2\pi})$ & 0.893823 & 1.08303
                    & 1.25109 & 1.40370 & 1.54440 & 1.67556 \\
$\eta_2$ & & & 1.17325 & 1.58860 & 1.95105 & 2.27645 & 2.57407 &
                                                                 2.84986 \\
$\eta_3$ & & & & & 2.17856 & 2.70016 & 3.17251 & 3.60715\\
$\eta_4$ & & & & & & & 3.36929 & 3.97952\\
\hline
\end{tabular}
\caption{Distinct parameters of the spin-chain Hamiltonian. The full
  set can be obtained using $\eta_{N-i}=\eta_i$. For $N
\leq3$ the coefficients are exact. For $N=4$ the
  coefficients may be obtained analytically \cite{Levinsen2015},
  however the resulting expressions are cumbersome and are not
  repeated here. }
\label{tab:eta}
\end{minipage} 
\end{table}

\begin{figure*}[h]
\centering
\vskip 0 pt
\includegraphics[width=.47\columnwidth,clip]{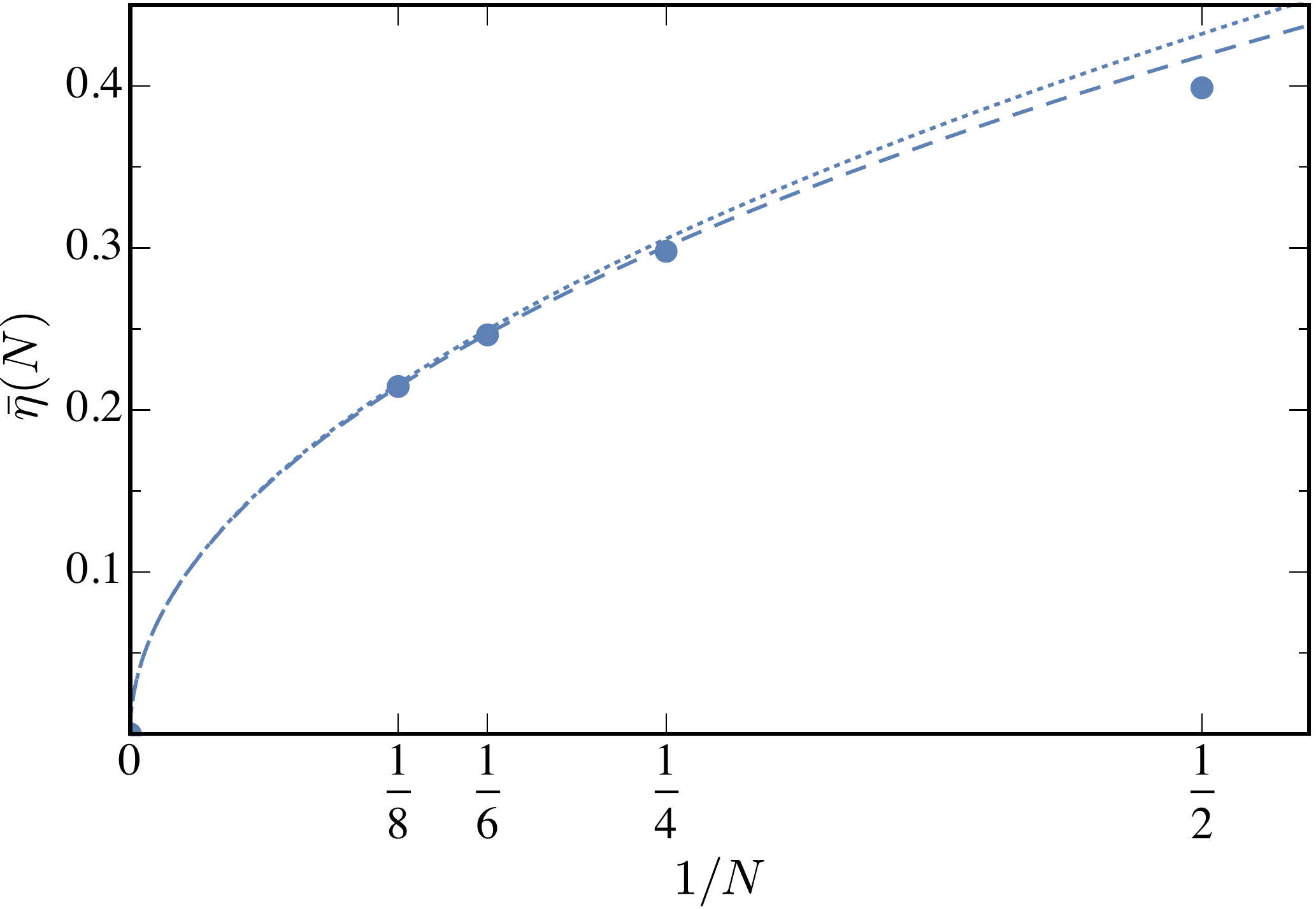}
\caption{$\bar\eta(N)$ as a function of $1/N$. For $N=2,4,6,8$ we use the
  exact result, indicated by dots, while for larger $N$ we approximate it by the dashed
  line, as outlined in the text [see Eq.\ \eqref{etabarlargeN}]. The dotted line corresponds to
  $\bar\eta(N)$ calculated from the large $N$ limit,
  Eq.~\eqref{eq:etainf}.
}
\label{fig:etabar}
\end{figure*}

In the manuscript, instead of using the numerically exact coefficients
in Table \ref{tab:eta} we use an approximate model derived in
Ref.~\cite{Levinsen2015}, where $\eta_i\equiv i(N-i) \bar\eta(N)$. Note that $\bar\eta(N)$ only sets the absolute
value of the energy and does not affect the eigenstates or the energy
ratios for each $N$.  We
fix $\bar\eta(N)$ as following: At $\phi=\pi$, corresponding to
infinite interspecies interactions, we know the exact ground state
which is fully phase separated. Thus, we may evaluate the ground state
energy to find
\begin{align}
  E_0=\varepsilon_0-\frac4{g_{\nu\nu}}\left[-\eta_{N/2}+\sum_{i=1}^{N-1}\eta_i\right].
\end{align}
Requiring that we reproduce the exact ground state energy at $\phi=\pi$
thus results in the simple expression for $\bar\eta(N)$:
\begin{align}
\bar\eta(N)=\frac{\sum_{i=1}^{N-1}\eta_i-\eta_{N/2}}{N^3/6-N^2/4-N/6},
\end{align}
with $\eta_i$ those of Table \ref{tab:eta}. This procedure yields
$\bar\eta(2)=\sqrt{1/2\pi}$, $\bar\eta(4)=0.297941$,
$\bar\eta(6)=0.246318$, and $\bar\eta(8)=0.214440$. For larger $N$ the
exact coefficients are not known, and we instead extrapolate
$\bar\eta$ towards the thermodynamic limit $N\to\infty$. In this
limit, the domain wall may be ignored, and the energy to leading
non-trivial order is that of a Lieb-Liniger gas, Eq.~\eqref{LLcontact} in the main text:
\begin{align} 
E_{\rm LL} = \frac{N^2}{2} - \frac{128\sqrt{2}}{45\pi^2} \frac{N^{5/2}}{g_{\nu\nu}}.
\label{eq:LLsup}
\end{align}
This result was obtained in Ref.~\cite{Astrakharchik2005} by applying
the local density approximation to the uniform space Bethe ansatz
result. Hence, in the limit $N\to\infty$, the function $\bar\eta(N)$
is, to leading order,
\begin{align}
\bar\eta_{\rm LL}(N)=\frac{64\sqrt{2}}{15\pi^2}\frac1{\sqrt{N}}.
\label{eq:etainf}
\end{align}
To go to finite $N$,
we choose the simplest possible extrapolation, requiring that
the ${\cal O}(N^{3/2})$ correction to Eq.~\eqref{eq:etainf} produces
the correct $\bar\eta(8)$. The result for $N\geq8$ is
\begin{align}
\bar\eta(N)\approx \bar \eta_{\rm LL}(N)
\left(1-0.063343/N\right),
\label{etabarlargeN}
\end{align}
which is shown in Fig.~\ref{fig:etabar}.

\section{Accuracy of the strong-coupling ansatz}

\begin{figure}[h]
\centering
\vskip 0 pt
\includegraphics[width=.5\columnwidth,clip]{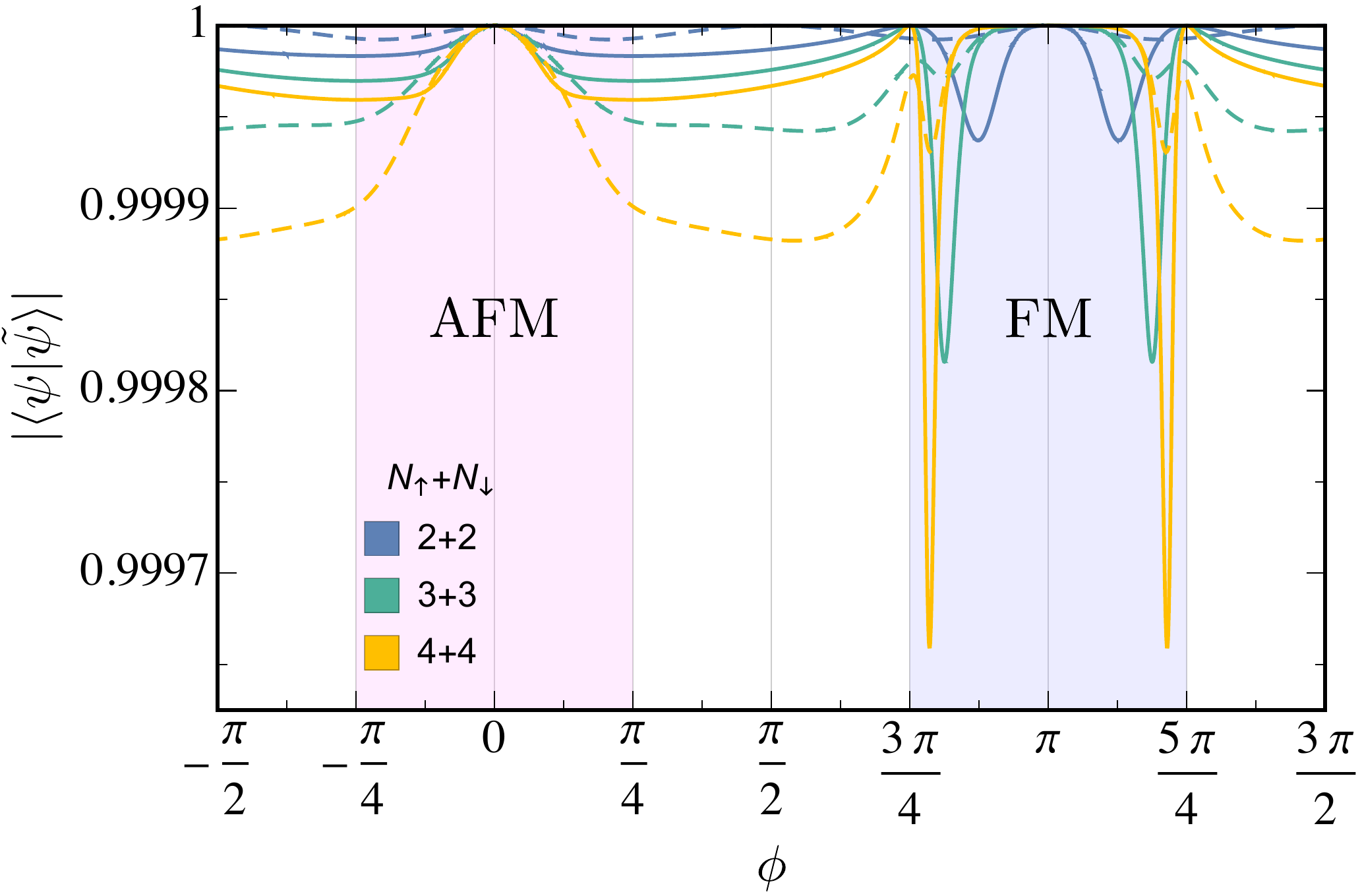}
\caption{Wavefunction overlaps between exact and approximate ground (solid) and
  first excited (dashed) states for $N_\up=N_\down=2$, 3, and 4
  particles.
}
\label{fig:overlaps}
\end{figure}

The high accuracy of the strong-coupling ansatz developed in
Ref.~\cite{Levinsen2015} and used here to determine the spin-exchange
coefficients beyond the few-body limit may be confirmed by computing
the overlap between the eigenstates $|\tilde \psi\rangle$ of the
approximate model and the eigenstates $|\psi\rangle$ obtained in the
exact solution of the problem. This we have computed for balanced
systems containing at most 8 particles, analogously to what was done
in our earlier work \cite{Levinsen2015}. As may be seen in Fig.\
\ref{fig:overlaps}, the wavefunction overlaps for both the ground and first excited
state (in the upper branch) remain extremely close to 1 for all ratios
of intra- and interspecies interaction strengths. The largest
deviations (of at most a few parts in $10^4$) are found near the
boundaries of the FM region.

\section{Spin gaps}
The stability of the system in the presence of a small inter-conversion coupling may be analyzed in terms of the ``spin gap",
i.e., the
energy difference between the ground state of a system of $N$
particles, and the configuration in which one of the spins is
flipped.
In the FM phase ($3\pi/4<\phi<5\pi/4$), the ground state
corresponds to a fully polarized gas of $N$ identical bosons, and we thus define the FM spin gap to be  
$\Delta E_{\rm FM}=E_0(N-1,1)-E_0(N,0)$, where $E_0(N_\up,N_\down)$ is the ground state energy of a gas composed of $N_\up$ spin-up particles and $N_\down$ spin-down ones.
For all other values of $\phi$, including those in the AFM phase, the ground state corresponds to a balanced system with $N_\up=N_\down$. Thus  
the spin gap in this region is instead defined as  $\Delta E_{\rm AFM}=E_0(N/2-1,N/2+1)-E_0(N/2,N/2)$.

At the classical Ising points $\phi=0,\pi$, we know the ground states
exactly and thus we also know the spin gap, which at both points is
$\Delta E=2\bar\eta(N)(N-1)/g$. This follows from the observation that
in both cases the energy cost associated with flipping a spin is
smallest at the trap edge.  We then apply perturbation theory to
estimate the behaviour of the spin gap around these points. In the AFM
regime, we have for $|\phi| \ll 1$,
\begin{eqnarray}
\label{AFMspinGap1}
\Delta E_{\rm AFM} & \simeq & 2 \frac{\bar\eta(N)}{g}
\left[ N-1 - (N+3)\phi^2 \right]\\
& \stackrel{N\to\infty}{\longrightarrow} & \frac{2N}{g}\bar\eta_{\rm LL}(N)(1-\phi^2),
\label{AFMspinGap2}
\end{eqnarray}
where in Eq.~\eqref{AFMspinGap1} we have kept the leading order correction
in $N$.
We display the analytic result,  Eq.~\eqref{AFMspinGap2}, in
Fig.~\ref{fig:degeneracyAndCorrelationsAndSpinGaps}(c)  to demonstrate
the manner in which the AFM spin gap approaches the thermodynamic limit.

We may perform a similar perturbative expansion around the FM Ising point, but in this case we find that we can describe the whole FM region for large $N$ with the following analytic form:
\begin{align} 
\Delta E_{\rm FM} 
\simeq 2 (N-1)\frac{ \bar\eta_{\rm LL}(N)}{g}
\sqrt{\cos^2\phi-\sin^2 \phi},
\label{FMspinGap}
\end{align}
as shown in Fig.~\ref{fig:degeneracyAndCorrelationsAndSpinGaps}(c).

\end{document}